\documentclass[12pt]{article}
\usepackage{ae}
\usepackage{graphicx}
\usepackage{psfrag}
\usepackage{epsfig}
\usepackage{rotating}
\usepackage{amssymb,amsmath}
\usepackage[T1]{fontenc}
\usepackage[latin1]{inputenc}
\usepackage{color}
\usepackage{subfigure}
\usepackage{slashed}

\usepackage{slashed}

\def\bfl{\begin{flushleft}}
\def\efl{\end{flushleft}}

\begin{document}
\begin{titlepage}


\vspace*{0.2cm}
\begin{center}
{\Large Methods for deriving functional equations  \\
\vspace{2mm}
  for Feynman integrals ${}^*$}
\\[2 cm]

\medskip

\author{
{\sc O.~V.~Tarasov}\thanks{On leave of absence from
Joint Institute for Nuclear Research,
141980 Dubna (Moscow Region), Russia.}
\\
}
{\bf  O.V.~Tarasov}\\[0.9 cm]
 {
    Joint Institute for Nuclear Research,\\
      141980 Dubna, Russian Federation \\
 \medskip
     {\it E-mail}: {\tt otarasov@jinr.ru}}\\

\end{center}

\vspace*{1.0cm}

\begin{abstract}
We present short review of two methods 
 for obtaining functional equations
for Feynman integrals.  
Application of these methods for finding  functional 
equations for  one- and two- loop integrals  
described in detail.
It is shown that with the aid of functional equations
Feynman integrals in general kinematics
can be expressed in terms  of  simpler integrals.  
Similarities between  functional equations for Feynman  integrals
and addition theorem for Abel integrals are shortly
discussed.
\medskip

\medskip

\vspace{2cm}

\medskip


\end{abstract}
\hrulefill
\hrule width 16cm
\bfl
${}^*$\footnotesize{To appear in the proceedings of the 4th Computational Particle Physics Workshop, October 2016, Hayama, Japan.}
\efl

\end{titlepage}


\tableofcontents


\section{Introduction}
Functional equations (FE) for Feynman integrals for the first time
were proposed in Ref. \cite{Tarasov:2008hw}. Details of derivation of
such equations for one-loop integrals  were described in Refs. \cite{Tarasov:2011zz},
\cite{Kniehl:2009pv}.
In Ref.\cite{Tarasov:2015wcd} a new method for derivation of
functional equations was proposed. This method is based on
finding algebraic relations between products of propagators.
By integrating different combinations of these algebraic relations
one can  easily  obtain functional equations for multiloop integrals.
Our functional relations can be used for expressing
integrals with nontrivial kinematical dependence in terms of
much simpler integrals. Also functional equations
can be exploited for analytic continuation of Feynman integrals into
different kinematical domains.
In the next sections we will explain how to derive FE and will
present several illustrative examples.

\section{Derivation of FE
from recurrence relations}

The method proposed in Ref.\cite{Tarasov:2008hw} is based on the use of
recurrence relations between integrals. The most general recurrence relation
for Feynman integrals can be  written as
\begin{equation}
 \sum_j Q_j I_{j,n}=\sum_{k,r<n}R_{k,r}I_{k,r},
\end{equation}
where $Q_j,R_k$ are polynomials in masses, scalar
products of external momenta,  space-time dimension
{$ d$}, and powers of propagators.
{ $I_{k,r} $} - are integrals 
with 
$r$ external lines. The index $k$ just labels different integrals in
a recurrence relation.
In recurrence relations some integrals are more complicated than the 
others. The complexity of the integral depends on the
number of their arguments and number of loops. The main idea of deriving functional
equation from a recurrence relation can be formulated as follows.
%
By choosing kinematical variables, masses, indices
of propagators remove most complicated
integrals from the relation, i.e. impose conditions :
\begin{equation}
Q_j=0,
\end{equation}
{
keeping  at least some other coefficients
{ $ R_k \neq 0 $}}.

To demonstrate how this idea works in practice let's consider
 one-loop   $n$-point integrals  { $I_n^{(d)}$}.
These integrals 
satisfy the so-called generalized recurrence relations~
\cite{Tarasov:1996br}:
\begin{equation}
{G_{n-1}} \nu_j{\bf j^+} I^{(d+2)}_n -({\partial_j
\Delta_n)}I^{(d)}_n =
\sum_{k=1}^{n}
 (\partial_j \partial_k \Delta_n)
 {\bf  k^-}  I^{(d)}_n,
\label{reduceJandDtod}
\end{equation}
where  {${\bf j^{\pm }}$}  shifts indices
{ $\nu_j \to \nu_{j } \pm 1$},
$
 \partial_j \equiv \frac{\partial }{ \partial m_j^2},
$
\begin{equation}
{ G_{n-1}}= -2^n \left|
\begin{array}{cccc}
  p_1p_1  & p_1p_2  &\ldots & p_1p_{n-1} \\
  \vdots  & \vdots  &\ddots & \vdots \\
  p_1p_{n-1}  & p_2p_{n-1}  &\ldots & p_{n-1}p_{n-1}
\end{array}
\right|,
\label{Gn}
\end{equation}
\begin{equation}
{ \Delta_n}=  \left|
\begin{array}{cccc}
Y_{11}  & Y_{12}  &\ldots & Y_{1n} \\
\vdots  & \vdots  &\ddots & \vdots \\
Y_{1n}  & Y_{2n}  &\ldots & Y_{nn}
\end{array}
         \right|,
~~~~~
Y_{ij}=m_i^2+m_j^2- { p_{ij}},~~~~~~~~~~~~~
{{ \bf \it  p_{ij}=(p_i-p_j)^2}}.
\end{equation}

\subsection{One-loop propagator type integral}
Setting { $n=3$, $j=1$}, {$m_3^2=0$} in Eq.(\ref{reduceJandDtod}) leads to
an expression connecting integrals $I_3^{(d+2)}$, $I_3^{(d)}$ and $I_2^{(d)}$.
In order to get rid of integrals $I_3^{(d+2)}$, $I_3^{(d)}$  we must
impose conditions:
\begin{equation}
G_2=0,~~~~~~~~~~\Delta_3=0. 
\end{equation}
Solving this system for $p_{13}$, $p_{23}$ yields FE for the
integral $I_2^{(d)}$:
\begin{eqnarray*}
\label{mm3zero}
I_2^{(d)}({m_1^2,m_2^2,p_{12}}) &=&
\frac{p_{12}+m^2_1-m^2_2-\alpha_{12}}{2p_{12}}
~I_2^{(d)}({m_1^2,0,s_{13}})
\nonumber
\\
&+&
\frac{p_{12}-m^2_1+m^2_2+\alpha_{12}}{2p_{12}}
~I_2^{(d)}({0,m_2^2,s_{23}}),
\end{eqnarray*}
where
\begin{eqnarray*}
&&s_{13}=\frac{\Delta_{12}+2p_{12}m_1^2 
-(p_{12}+m_1^2-m_2^2)\alpha_{12}}
{2p_{12}},
\nonumber
\\
&&s_{23}=\frac{ \Delta_{12} +2p_{12} m_2^2
+(p_{12}-m_1^2+m_2^2) \alpha_{12} }
{2p_{12}},
\\
&&\alpha_{12}=\pm \sqrt{\Delta_{12}}.
\end{eqnarray*}
\[
\Delta_{ij}=p_{ij}^2+m_i^4+m_j^4-2p_{ij}m_i^2-2p_{ij}m_j^2-2m_i^2m_j^2.
\]
This equation represent  integral with arbitrary masses $m_1$, $m_2$
and arbitrary external momentum in terms of integrals with one propagator massless.
 $I_2^{(d)}$ integrals with one massless propagator can be expressed
 in terms of Gauss' hypergeometric function $_2F_1$
 \cite{Bollini:1972bi},
 \cite{Boos:1990rg}.
 Evaluation of such integrals is much easier than evaluation of
 integrals with two masses.

\subsection{FE for one-loop vertex type integral}

At { $n=4$, $j=1$, $m_4=0$}, in order to get rid of  integrals
$I_4^{(d)}$, $I_4^{(d+2)}$ and also one vertex integral $I_3^{(d)}$ 
in Eq.(\ref{reduceJandDtod}), we impose conditions
\begin{equation}
G_3=0, ~~~~\partial_1 \Delta_4=0,~~~~\partial_1 \partial_2 \Delta_4=0.
\end{equation}
This system of equations depends on 10 variables
 $p_{12},p_{13},p_{14},p_{23},p_{24},p_{34}$,
 $m_1^2,m_2^2,m_3^2,m_4^2$ and it can be solved 
by excluding, for example,  $p_{14}$, $p_{24}$ and $p_{34}$.
Substituting solution  for $p_{14}$, $p_{24}$, $p_{34}$ into
Eq.(\ref{reduceJandDtod}) leads to the following FE:
\begin{eqnarray}
&&I_3^{(d)}({ m_1^2,m_2^2,m_3^2,s_{23},s_{13},s_{12}})=
\nonumber \\
&&~~
\frac{ s_{13}+m_3^2-m_1^2+\alpha_{13} }{2s_{13}}
I_3^{(d)}({ m_2^2,m_3^2,0, 
   s_{34}^{(13)},
   s_{24}(m_1^2,m_3^2,s_{23},s_{13},s_{12}),  s_{23}})
\nonumber \\
&&~+
\frac{ s_{13}-m_3^2+m_1^2-\alpha_{13} }{2s_{13}}
 I_3^{(d)}({ m_1^2,m_2^2,0,
          s_{24}(m_1^2,m_3^2,s_{23},s_{13},s_{12}),
          s_{14}^{(13)},s_{12}}),
\label{mm4zero}
\end{eqnarray}
where
\begin{eqnarray}
\label{solu_mm4zero}
p_{14}&=&s_{14}^{(13)},
\nonumber 
\\
p_{24}&=&s_{24}(m_1^2,m_3^2,p_{23},p_{13},p_{12} )
\nonumber
\\
&=&\frac{ (p_{12}+p_{23}-m_1^2-m_3^2)p_{13}
+(p_{12}-p_{23}-m_1^2+m_3^2)(m_3^2-m_1^2 +\alpha_{13})}{2p_{13}},
\nonumber \\
 p_{34}&=&s_{34}^{(13)},
\end{eqnarray}
and
\begin{eqnarray}
s_{14}^{(ij)}&=&
 \frac{\Delta_{ij}+2m_i^2 p_{ij}-(p_{ij}+m_i^2-m_j^2)\alpha_{ij} }
 {2p_{ij}},
\nonumber
\\
s_{34}^{(ij)}&=& \frac{\Delta_{ij}+2m_j^2 p_{ij}
+(p_{ij}+m_j^2-m_i^2) \alpha_{ij} }{2p_{ij}},
\nonumber
\\
\alpha_{ij}&=&\pm\sigma(p_{ij}-m_i^2+m_j^2) \sqrt{\Delta_{ij}}.
\end{eqnarray}
Again as it was for  integral { $I_2^{(d)}$} integral
{ $I_3^{(d)}$} with arbitrary arguments can be expressed in terms
of integrals with at least one propagator massless. By applying 
formula (\ref{mm4zero}) to integrals on the right-hand side of this equation
several times one can express { $I_3^{(d)}$} with arbitrary arguments
in terms of integrals with two massless propagators
and one line with massive propagator. Also one of kinematical
arguments in these integrals will be zero and other arguments 
will be functions of initial kinematical arguments and masses.
This integral can be expressed in terms of Appell hypergeometric
function $F_1$ and  Gauss' hypergeometric function $_2F_1$ 
\cite{Tarasov:2008hw}.

\section{Derivation of FE by using  algebraic relations for propagators}

Analyzing  FE for one-loop Feynman integrals
one can observe that integrands are
rather similar and differ only by one propagator. In the relation
for one-loop propagator type integral integrands are 
\begin{eqnarray*}
\frac{1}{D_1 D_2},~~\frac{1}{D_0 D_2}, \frac{1}{D_1 D_0},
\end{eqnarray*}
and integrands for the one-loop vertex type integrals are
\begin{eqnarray*}
\frac{1}{D_1 D_2D_3},~~\frac{1}{D_0 D_2D_3}, \frac{1}{D_1 D_0 D_3},
\frac{1}{D_1 D_2D_0}
\end{eqnarray*}
where
\begin{eqnarray*}
&&D_0 = (k_1-p_0)^2-m_0^2+i\epsilon,~~~D_1 = (k_1-p_1)^2-m_1^2+i\epsilon,~~~
\\
&&D_2 = (k_1-p_2)^2-m_2^2+i\epsilon,~~~D_3 =
(k_1-p_3)^2-m_3^2+i\epsilon,
\end{eqnarray*}
and $i\epsilon$ is traditional causal prescription.
Here $k_1$ is integration momentum and  on scalar products of $p_1$,$p_2$,...
some restrictions, like  $G_n=0$, are imposed. Such restrictions
effectively lead to linear dependence of these vectors.

\subsection{Algebraic relations for products of propagators}

One can raise the question:
would it be possible to find algebraic 
relations for products of propagators and derive 
FE from such  relations? From the explicit FE for one-loop
propagator integral we can try to find relation of the form:
\begin{equation}
\frac{1}{D_1 D_2}=\frac{x_1}{D_0 D_2}+ \frac{x_2}{D_1 D_0}
\end{equation}
where we assume that
{ $x_1,x_2$, $p_{0}$, $p_{1}$ and $p_{2}$}
are independent of { $k_1$}.
Putting all terms over the common denominator we get
\begin{equation}
D_0=x_1D_1 + x_2D_2.
\label{D0_equatio}
\end{equation}
By differentiating (\ref{D0_equatio}) with respect to $k_1$ we obtain
two  equations:
\begin{equation}
x_1+x_2 = 1, ~~~~~~p_0 = x_1p_1 + x_2p_2.
\label{equs_by_diff}
\end{equation}
Taking into account these relationships we get from (\ref{D0_equatio})
an additional equation:
\begin{equation}
x_1x_2s_{12} + m_0^2-x_1m_1^2 -x_2m_2^2=0,
\label{equ_from_D0_equatio}
\end{equation}
where $s_{12}=(p_1-p_2)^2$.
From Eqs.(\ref{equs_by_diff}), (\ref{equ_from_D0_equatio}) one get
expressions for $x_1$,$x_2$:
{
\begin{eqnarray}
&&
x_1 = \frac{m_2^2-m_1^2+s_{12}}{2s_{12}}
-\frac{\sqrt{\Lambda_2+4s_{12}m_0^2}}{2 s_{12}},
\nonumber \\
&&
x_2 =\frac{m_1^2-m_2^2+s_{12}}{2s_{12}} 
+\frac{\sqrt{\Lambda_2+4s_{12}m_0^2}}{2 s_{12}},
\label{solution_I2}
\end{eqnarray}
}
 and
\begin{eqnarray*}
\Lambda_2=s_{12}^2+m_1^4+m_2^4-2s_{12}(m_1^2+m_2^2)-2m_1^2m_2^2.
\end{eqnarray*}
Integration of the obtained algebraic relation w.r.t. { $k_1$}
gives the following FE:
{
\begin{eqnarray*}
I_2^{(d)}({ m_1^2,m_2^2,s_{12}}) =
\frac{s_{12}+m_1^2-m_2^2 + \lambda}
 {2s_{12}}
~I_2^{(d)}({ m_1^2,m_0^2,s_{13}(m_1^2,m_2^2,m_0^2,s_{12})})
\nonumber 
\\
 +\frac{s_{12}-m_1^2+m_2^2 - \lambda}
 {2s_{12}}
 ~I_2^{(d)}({ m_2^2,m_0^2,s_{23}(m_1^2,m_2^2,m_0^2,s_{12})}),
\end{eqnarray*}
}
where
{
\begin{eqnarray*}
&&s_{13} = \frac{\Lambda_2+2 s_{12}(m_1^2+m_0^2)}{2 s_{12}}
+\frac{m_1^2-m_2^2+s_{12}}{2s_{12}} \lambda
\nonumber 
\\
&&
s_{23} =
 \frac{\Lambda_2
+2s_{12}(m_2^2+m_0^2)}
{2s_{12}} +\frac{m_1^2-m_2^2-s_{12}}{2s_{12}}\lambda. 
\\
&&
\lambda=\sqrt{\Lambda_2+4s_{12}m_0^2}
\label{S13_S23}
\end{eqnarray*}
}
Parameter { $m_0$} is arbitrary and can be taken at will.
The same equation was obtained from recurrence relations
by imposing conditions on Gram determinants. 

Similar to the relation with two propagators one can find
relation for  three propagartors:
\begin{eqnarray*}
\frac{1}{D_1 D_2 D_3}= \frac{x_1}{D_2 D_3 D_0} + 
\frac{x_2}{D_1D_3D_0} +\frac{x_3}{D_1D_2D_0} .
\label{1loop_3props}
\end{eqnarray*}
Here $p_0$, $p_1$, $p_2$ and $p_3$ are 
external momenta assumed to be independent on $k_1$.
Momentum $k_1$ will be integration momentum and we assume also
that $x_i$ do not depend on $k_1$.
Multiplying both sides of equation (\ref{1loop_3props}) by
the product $D_1D_2D_3D_0$ we get
\begin{equation}
D_0=x_1D_1 + x_2D_2 + x_3 D_3.
\label{D0_three_points}
\end{equation}
Differentiation of this equation with respect to $k_1$ gives
two equations:
\begin{equation}
x_1+x_2+x_3=1,~~~~p_0=x_1 p_1+ x_2p_2 +x_3p_3.
\end{equation}
Taking into account these relationships from Eq. (\ref{D0_three_points}) 
we obtain:
\begin{equation}
x_2x_3p_{23}+ x_1 x_3 p_{13} +
x_1x_2p_{12}-x_1m_1^2-x_2m_2^2-x_3m_3^2+m_0^2=0. 
\end{equation}
This system has the following solution
{
\begin{eqnarray*}
x_1 = 1-\alpha-x_2,~~~~~x_3 = \alpha, 
\end{eqnarray*}
}
where $\alpha$ is solution of the quadratic equation
{
\begin{eqnarray*}
&&\alpha^2 p_{13}
+[m_3^2-m_1^2-p_{13} +x_{2}(p_{13}+p_{12}-p_{23})] \alpha
\\ 
&&+m_1^2-m_0^2+(m_2^2 -m_1^2 - p_{12}+p_{12}x_{2})x_{2}=0.
\end{eqnarray*}
}
Solution depends on 2 arbitrary parameters: $m_0$, $x_{2}$. 
By integrating the obtained relation we get the same
FE as it was given before.

As was shown in Ref. \cite{Tarasov:2015wcd} one can  also
derive algebraic relations for products of any number of
propagators.

 Functional relations for Feynman integrals
with integrands being rational functions of scalar products 
of different momenta to some extent resemble
{\it Abel's addition theorem} 
\cite{markushevich1992introduction},
\cite{Barnum:1910},
\cite{vilenkin1992representation}. 
It would be useful to find similarities between Feynman integrals and
Abel integrals.

\subsection{Some remarks about Abel integrals}
In this subsection we will present some facts about Abelian integrals
that may be useful in  investigation of functional equations 
for Feynman integrals. 
Abelian integral is an integral in the complex plane of the form
{
\begin{eqnarray*}
\int_{z_0}^{z} R(x,w)dx,
\end{eqnarray*}
}
where $R(x,w)$ is an arbitrary {\it rational} function of the two variables
$x$ and $w$. These variables are related by the equation
{
\begin{eqnarray*}
F(x,w)=0,
\end{eqnarray*}
}
where $F(x,w)$ is an irreducible polynomial in $w$,
{
\begin{eqnarray*}
F(x,w) \equiv \phi_n(x)w^n + ...+ \phi_1(x)w + \phi_0(x),
\end{eqnarray*}
}
whose coefficients $\phi_j(x)$, $j=0,1,...n$ are rational
functions of $x$.
Abelian integrals are natural generalizations of elliptic integrals, which
arise when
{
\begin{eqnarray*}
F(x,w)=w^2-P(x),
\end{eqnarray*}
}
where $P(x)$ is a polynomial of degree 3 and 4. If degree of the 
polynomial is greater than 4 then we have {\it hyperelliptic integral}.
For Abel integrals  an important theorem was proven.
\\
Let $C$ and $C'$ be plane curves given by the equations
{
\begin{eqnarray*}
&& C:~~F(x,y)=0,\\
&& C':~~\phi(x,y)=0.\
\end {eqnarray*}
}
These curves have $n$ points of intersections {
$(x_1,y_1)$,{\ldots} $(x_n,y_n)$}, where $n$ is the product of
degrees of $C$ and $C'$. Let $R(x,y)$ be a rational function
of $x$ and $y$ where $y$ is defined as a function of $x$ by the
relation { $F(x,y)=0$}. Consider the sum
\begin{equation}
I=\sum_{i=1}^{n} \int_{x_0,y_0}^{x_i,y_i} R(x,y) dx.
\end{equation}
Integrals being taken from a fixed point to the $n$ points of
intersections of $C$ and $C'$. If some of the coefficients $a_1$,$a_2$,{\ldots} ,$a_k$
of $\phi(x,y)$ are regarded as continuous variables, the points
$(x_i,y_i)$ will vary continuously and hence  { $I$}
will be a function, whose form is to be determined, of the variable
coefficients $a_1$,$a_2$,{\ldots},$a_k$.\\
\noindent
{\it Abel's theorem:}\\
The partial derivatives of the sum { $I$} , with respect to any of the
coefficients of the variable curve { $\phi(x,y)=0$},
is a {\it rational} function of the coefficients and hence
{ $I$} is equal to a {\it rational} function of the 
coefficients of { $\phi(x,y)=0$} , plus a finite number
of logarithms or arc tangents of such rational functions.
What is very  important - integrals themselves can be rather complicated 
transcendental functions but their sum can  be simple.
\\
Example: Elliptic integral of the second type \cite{Barnum:1910}:
{
\begin{eqnarray*}
E(k,x)=\int_0^{x}\frac{(1-k^2x)dx}{\sqrt{x(1-x)(1-k^2x)}}.
\end{eqnarray*}
}
We take as $C$ and $C'$ 
\begin{equation}
C:~~~y^2=x(1-x)(1-k^2x),~~~~~~~~~~~~~~~~~~~~~~C':~~~y=ax+b.
\end{equation}
The elimination of $y$ between two equations will give us  as the
abscissae $x_1,x_2,x_3$ of the points of intersection the three roots 
of the equation:
\begin{equation}
\phi(x)=k^2x^3 -(1+k^2+a^2)x^2+(1-2ab)x-b^2=0.
\label{three_roots}
\end{equation}
The corresponding sum will be
\begin{eqnarray*}
I(a,b)=\int_0^{x_1}R(x,y)dx +\int_0^{x_2}R(x,y)dx +\int_{1/k^2}^{x_3}R(x,y)dx,
\end{eqnarray*}
where
\begin{equation}
R(x,y)=\frac{1-k^2x}{y}.
\end{equation}
Abel's theorem gives addition formula:
\begin{equation}
\int_0^{x_1}R(x)dx +\int_0^{x_2}R(x)dx +\int_{1/k^2}^{x_3}R(x)dx = -2a
+ \kappa,
\label{equ_for_elliptic}
\end{equation}
where $\kappa$ is an arbitrary constant.

One can see that  FE for Feynman integrals
and relationship   (\ref{equ_for_elliptic}) have some common features.
Arguments of Feynman integrals and arguments of functions on the left-hand side
of Eq. (\ref{equ_for_elliptic}) are determined from algebraic
equations and integrands in both cases are rational functions
of some algebraic functions.

One can find FE for Feynman integrals following closely derivation
of relationships for usual algebraic integrals. 
Deriving relations for propagators we used orthogonality condition
$G_n=0$.
In fact it is not needed to assume such a relation. For example, to fix parameters 
in algebraic relations for products of two propagators 
{
\begin{equation}
R_2(k_1,p_1,p_2,m_1^2,m_2^2,m_0^2)=\frac{1}{D_1D_2}- \frac{x_1}{D_2D_0}
-\frac{x_2}{D_1D_0}=0,
\label{prod2props}
\end{equation}
}
instead of $G_n=0$ we can impose conditions
\begin{equation}
\frac{\partial x_1}{\partial k_{1\mu}}=\frac{\partial x_2}{\partial
k_{1\mu}}=0,
\label{x1x2_k1_indep}
\end{equation}
having in mind that only $D_j$  factors will depend on momentum 
$k_1$.
Multiplying both sides of Eq. (\ref{prod2props}) by $D_0D_1D_2$ we get
{
\begin{eqnarray}
&&D_0-x_1 D_1-x_2 D_2 
\nonumber
\\
&&
=(1-x_1-x_2)k_1^2 + 2x_1k_1p_1 + 2x_2k_1p_2
\nonumber
\\
&&
+x_1m_1^2+x_2m_2^2-x_1p_1^2 - x_2p_2^2 - 2k_1p_0 - m_0^2+p_0^2.
\label{rel4props}
\end{eqnarray}
}

Differentiating this relation w.r.t. $k_{1\mu}$, contracting with
{ $k_1,p_1,p_2,p_0$} and taking into account
(\ref{x1x2_k1_indep}) gives several equations:
{
\begin{eqnarray}
&&       -2 x_1 (k_1^2 - k_1p_1) - 2x_2 (k_1^2 - k_1p_2) + 2 k_1^2 - 2
k_1p_0 =0,
\nonumber
\\
&&       2(1-x_1 - x_2)  k_1p_1 + 2 x_1 p_1^2 + 2 x_2 p_1p_2 - 2 p_1p_0 =0,
\nonumber
\\
&&       2(1-x_1 - x_2)  k_1p_2 + 2 p_1p_2 x_1 + 2 x_2 p_2^2 - 2 p_2p_0=0,
\nonumber\\
&&       2(1-x_1 - x_2)  k_1p_0 + 2 x_1 p_1p_0 + 2 x_2 p_2p_0 - 2
p_0^2=0.
\end{eqnarray}
}
They can be used to express  {
$k_1p_0$, $p_1p_0$, $p_2p_0$, $x_1$, $x_2$} in terms of
 { 
$k_1^2$, $k_1p_1$, $k_1p_2$, $p_1^2$, $p_1p_2$, $p_2^2$}
considered to be independent variables. 
For example, we get:
\begin{equation}
k_1p_0=
x_1 k_1p_1-x_1 k_1p_2+k_1p_2+\frac{x_1}{2}(m_1^2-m_2^2-p_1^2+p_2^2)
+\frac12(m_2^2- p_2^2-m_0^2+p_0^2),
\end{equation}
and similar expressions for other scalar products of $p_0$.
Solution for { $x_1$,  $x_2$} is the same as
in Eq. (\ref{solution_I2}) and as a result we obtained the same relation
between products of two propagators.

Solution of the above system of equations is rather similar to
finding intersections of two plane curves considered in
Abel's theorem. Unfortunately extension of Abel's theorem for
integrals of algebraic functions of several variables is not an
easy task. In the case of functions of two variables some
results were obtained long time ago in Refs. \cite{picard1906theorie},
\cite{Forsyth1882}.

Similar to usual algebraic integrals of one variable we can 
construct various integrands out of
our different relationships 
for products of propagator.
These integrands will be rational functions in independent 
variables.
Integrations should be done over $d$ dimensional space.
Rational function must resemble integrands for Feynman integrands. 

For example,  multiplying $R_2({k_2},p_2,p_4,m_2^2,m_4^2,\widetilde{m}_0^2)$ by 
\begin{equation}
\frac{1}{  [(k_1-p_1)^2-m_1^2]^{\nu_1}
[(k_1-p_3)^2-m_3^2]^{\nu_3}[(k_1-k_2)^2-m_5^2]^{\nu_5}},
\end{equation}
and integrating this product 
with respect to $k_1$, $k_2$ leads to  the following FE:
\begin{eqnarray}
\int\!\frac{d^dk_1d^dk_2~~~~R_2({k_2},p_2,p_4,m_2^2,m_4^2,\widetilde{m}_0^2)}
{  [(k_1-p_1)^2-m_1^2]^{\nu_1}[(k_1-p_3)^2-m_3^2]^{\nu_3}
  [(k_1-k_2)^2-m_5^2]^{\nu_5}}\! =0,
\end{eqnarray}
Integrals in this equation correspond to the  diagram given in Fig.1.
\begin{center}
\scalebox{.65}{ \includegraphics{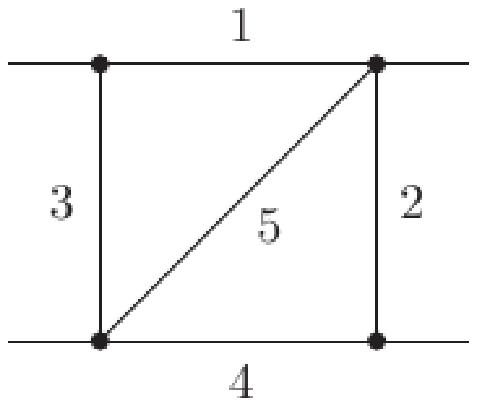}}
\end{center}
\begin{center}
{Fig.1. FE for the two-loop box integral.}\\
\end{center}

By  integrating product of the relationship
{
\begin{eqnarray*}
&&R_3(k_1,p_1,p_2,p_3,m_1^2,m_2^2,m_3^2,m_0^2)
\\
&&=\frac{1}{D_1D_2D_3}- \frac{x_1}{D_2D_3D_0}
-\frac{x_2}{D_1D_3D_0}-\frac{x_3}{D_1D_2 D_0}=0,
\end{eqnarray*}
}
and one loop-propagator integral
{
\begin{eqnarray*}
\int \int\frac{d^d k_1d^dk_2}{[(k_1-p_1)^2-m_1^2]^{\nu_1}[(k_1-k_2)^2-m_5^2]^{\nu_5}}
R_3({ k_2},p_2,p_3,p_4,m_2^2,m_3^2,m_4^2,m_0^2)=0,
\end{eqnarray*}
}
we obtain the FE for the integral corresponding to the following diagram
with arbitrary $\nu_1$, $\nu_5$, and arbitrary momenta and masses
\begin{center}
\scalebox{.65}{\includegraphics{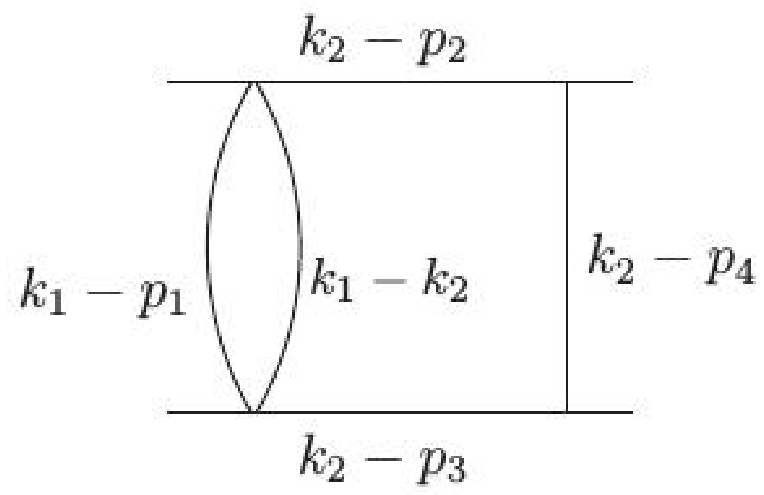}}
\end{center}
\begin{center}
{Fig.2. FE for the two-loop integral with propagator insertion.}\\
\end{center}
By integrating different products of $R_2$, $R_3$ and products of
different propagators with respect to selected momenta one can easily derive 
FE for various multiloop Feynman integrals.

\section{Concluding remarks}

We described two different methods for finding FE 
for Feynman integrals with any number of
loops and external legs.
FE reduce integrals with complicated kinematics
to simpler integrals.
FE can be used for analytic continuation of Feynman
integrals without knowing explicit analytic result.  
At the present time application of these methods for some two- and three- loop
integrals is in progress.
 Systematic investigation of  FE for Feynman integrals
based on algebraic geometry and group theory is needed.
Some improvements of these methods can be done
by exploiting known methods for algebraic integrals. 
The methods can be extended for finding
functional equations among hypergeometric as well as
holonomic functions.

\section{Acknowledgment}
I would like to thank organizers of the  CPP2016 for the
invitation and great workshop. My contribution to the  proceedings I dedicate
to the memory of Shimizu - Sensei.
I am also thankful to referee for useful suggestions taken into account 
in Section 3 of this paper.


\begin{thebibliography}{20}

\bibitem{Tarasov:2008hw}
Tarasov O V 2008
\newblock {New relationships between Feynman integrals}.
\newblock {\em Phys.Lett.}, B670:67--72

\bibitem{Tarasov:2011zz}
Tarasov O V  2011
\newblock {Functional equations for Feynman integrals}.
\newblock {\em Phys.Part.Nucl.Lett.}, 8:419--427

\bibitem{Kniehl:2009pv}
Kniehl B A and  Tarasov O V  2009
\newblock {Functional equations for one-loop master integrals for heavy-quark
  production and Bhabha scattering}.
\newblock {\em Nucl.Phys.}, B820:178--192

\bibitem{Tarasov:2015wcd}
Tarasov O V  2015
\newblock {Derivation of Functional Equations for Feynman Integrals from
  Algebraic Relations}.
\newblock {\em Preprint} hep-ph/1512.09024

\bibitem{Tarasov:1996br}
 Tarasov O V  1996
\newblock {Connection between Feynman integrals having different values of the
  space-time dimension}.
\newblock {\em Phys.Rev.}, D54:6479--6490.

\bibitem{Bollini:1972bi}
~~Bollini C G and  Giambiagi J J  1972
\newblock {Lowest order divergent graphs in nu-dimensional space}.
\newblock {\em Phys. Lett.}, B40:566--568

\bibitem{Boos:1990rg}
Boos E E and Davydychev A I  1991
\newblock {A Method of evaluating massive Feynman integrals}.
\newblock {\em Theor.Math.Phys.}, 89:1052--1063

\bibitem{markushevich1992introduction}
~Markushevich A I 1992
\newblock {\em Introduction to the Classical Theory of Abelian Functions}.
\newblock Translations of mathematical monographs. American Mathematical
  Society


\bibitem{Barnum:1910}
~Barnum H H 1910 
\newblock Abel's theorem and the addition formulae for elliptic integrals.
\newblock {\em Annals of Mathematics}, 11(3):103--114

\bibitem{vilenkin1992representation}
~Vilenkin N J and  Klimyk A U 1992
\newblock {\em Representation of Lie Groups and Special Functions: Volume 3:
  Classical and Quantum Groups and Special Functions}.
\newblock Mathematics and its Applications. Springer Netherlands

\bibitem{picard1906theorie}
~Picard E and Simart G  1897,1900, 1904,1906
\newblock {\em Theorie des fonctions algebriques de deux variables
  independantes}, volume 1,2.

\bibitem{Forsyth1882}
~Forsyth A R 1882
\newblock On Abel's theorem and abelian functions.
\newblock {\em Proceedings of the Royal Society of London}, 34:288--291

\end{thebibliography}
\end{document}